\documentclass[pre,superscriptaddress,twocolumn,sort&compress,round,footinbib]{revtex4-1}

\usepackage{graphicx}
\usepackage{gensymb}
\usepackage{natbib}
\usepackage[dvipsnames]{xcolor}
\usepackage{hyperref}
\hypersetup{
    colorlinks = true,
    urlcolor   = blue,
    citecolor  = black,
}
\newcommand{\SImum}{\textrm{\textmu{}m}}
\definecolor{firebrick}{rgb}{0.7, 0.13, 0.13}

\begin{document}
\title[Role of liquid driving on clogging of suspensions]{Role of liquid driving on the clogging of constricted particle suspensions}

\author{Mathieu Souzy}
\email{mathieu.souzy@inrae.fr}
\affiliation{INRAE, Aix-Marseille Univ, UMR RECOVER, 13182 Aix-en-Provence, France}

\author{Alvaro Marin}
\email{a.marin@utwente.nl}

\affiliation{Physics of Fluids, University of Twente, The Netherlands}

\begin{abstract}
Forcing dense suspensions of non-cohesive particles through constrictions might either result in a continuous flow, an intermittent one, or indefinite interruption of flow, i.e., a clog.
While one of the most important (and obvious) controlling parameters in such a system is the neck-to-particle size ratio, the role of the liquid driving method is not so obvious. On the one hand, wide-spread volume-controlled systems result in pressure and local liquid velocity increases upon eventual clogs. On the other hand, pressure-controlled systems result in a decrease of the flow through the constriction when a clog is developed. 
The root of the question therefore lies on the role of interparticle liquid flow and hydrodynamic forces on both the formation and stability of an arch blocking the particle transport through a constriction.  
In this work, we experimentally analyse a suspension of non-cohesive particles in channels undergoing intermittent regimes, in which they are most sensitive to parametric changes. By exploring the statistical distribution of arrest times and of discharged particles, we surprisingly find that the transport of non-cohesive suspensions through constrictions actually follows a ``slower is faster'' principle under certain conditions.
\end{abstract}

\maketitle



\section{Introduction}
\label{sec:Intro}

The blockage of mass flow through constrictions is unfortunately a common issue in industrial processes, which is typically solved by simply substituting the clogged portion of the conduct, with the subsequent elevated costs. This phenomenon can manifest in different ways depending on the nature of the material being transported. For example, when colloidal suspensions are being passed through narrow passages, colloidal particles often feel an attraction towards the walls and to each other. Under such conditions, particles tend to accumulate in the vicinity of the walls and such a particle aggregate can grow until it blocks the channel \citep{Wyss2006,Delouche2021}. {In such circumstances, cohesive suspensions can also provoke clogs by successive deposition, $i.e.$ particles progressively depositing on the wall thus reducing the constriction size \citep{Agbangla2012, Delouche2020, Dersoir2015clogging, duru2015three}.}


Nonetheless, particles can also clog a constriction purely by mechanical forces, in the absence of cohesive forces towards the wall or towards other particles. This can easily occur when particles (typically larger than colloidal size) form a stable arch that blocks the outlet or the narrow channel, see Figure \ref{fig:setup}. This is a common way of blocking a passage, not only in granular systems \citep{Zuriguel2014}, but also in car traffic flow \citep{helbing2004physics}, people escaping in panic \citep{Garcimartin2016pedestrians} or in animal herds \citep{Garcimartin2015}. Although arching in suspensions has several analogies with such many-body systems, suspensions have the special feature of being a two-phase flow, in which both particles and liquid must pass through the constriction. This is a clear and crucial difference, but its implications are still unclear.  

Statistical analysis on many-body flow through constrictions   have identified very clear patterns \citep{Zuriguel2020} also in suspensions flows \citep{Marin2018,Souzy2020}, but the role of the liquid flow is far from being well understood. A fluidic system like the one shown in Figure \ref{fig:setup} can be driven by forcing a constant volume of fluid (volume-controlled), which will drag the solid particles along. When an arch is formed, the flow encounters high hydrodynamic resistance in the constriction and the local liquid velocity at the constriction will increase. On the contrary, if the system is driven by a hydrodynamic pressure difference, the liquid flow velocity will reduce at the constriction. Which situation will be more favourable for the mass transport through the constriction? in other words, does a higher local flow velocity at the constriction favour the particle transport or does it rather stabilize arches by pushing particles harder together? the answer is far from trivial. 

In order to answer this crucial question, in this paper we aim to elucidate the role of the liquid driving method in constricted flows of non-cohesive suspensions, in which the arch formation is entirely due to mechanical forces. Our method is based on reproducible experiments with significant amount of data from which we can obtain significant statistics on both arch formation and destruction. Since the parametric space is so large, we choose to perform in experiments in systems undergoing an intermittent flow regime, in which they are most sensitive to the control parameters. Using the same statistical tools as those employed in granular matter permits us also to make a direct comparison to those ``dry'' counterparts and have a deeper understanding on the role of liquid flow in the clogging of constricted suspensions by mechanical forces.

\section{Experiments}\label{exp}

\begin{figure*}
\centerline{\includegraphics[width=1\textwidth]{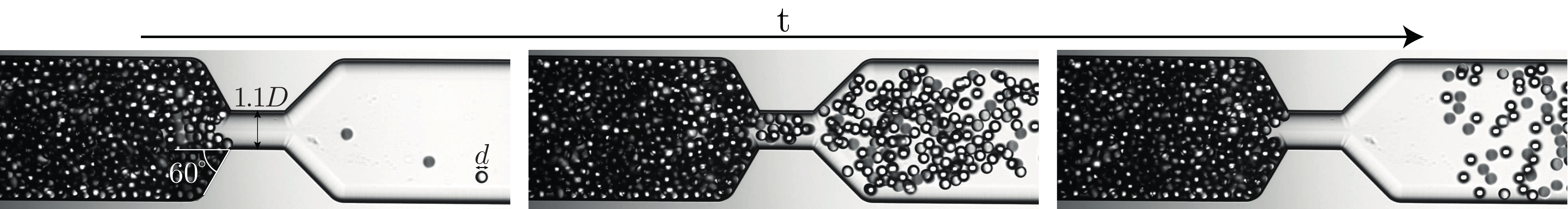}}
	\caption{Successive snapshots of a burst in a suspension of particles with a diameter $d$ that intermittently flows through a constriction having a neck width $1.1D$ and height $D=100~\SImum$ with $D/d=3.03$. From left to right: clog previous to a burst, burst, and clog after the burst. {The first and the last frame are separated by approximately 10 ms.}}
	\label{fig:setup}
\end{figure*}

Our experimental system is designed to flow in both pressure or volume-controlled driving in an intermittent regime, in which particles flow in bursts, interrupted by the formation of a particle arch that interrupts the flow. Our aim is to obtain (1) the average amount of particles per burst and (2) the probability distribution of time-lapses, the time passing between bursts. The particle flow is characterized experimentally using these statistical tools, which assess how the system responds to either pressure or liquid volume-rate changes. 

The experimental set-up is shown in Figure \ref{fig:setup}. The fluidic system consists of a transparent straight channel of borosilicate glass (isotropic wet etching, Micronit microfluidics) with a rectangular cross-section of $100 \times 400\,\SImum^2$ which reduces to an almost square cross-section of $100\times110\,\SImum^2$ to form the neck. The constriction is achieved by a linear narrowing of the channel with a half-angle of 60\degree. This specific design, similar of that used in previous studies \citep{Marin2018, Souzy2020}, forms a two-dimensional nozzle converging towards the neck. Particles and liquid have been carefully chosen to avoid buoyancy effects, particle aggregation and particle deposition at the micro-channel walls, an effect which has already been previously analysed \citep{Dressaire2017,Cejas2017, Delouche2020, Delouche2021, sendekie2016colloidal}. The suspension consists of monodisperse {\color{black} spherical} polystyrene particles of diameter $d=19.0, 21.7,\,\rm{and}\,33.0\,\SImum\,(\pm3\%$). Adopting the neck height $D=100\,\SImum$ as the characteristic length scale, these correspond to neck-to-particle ratios $D/d= 5.26, 4.61,\,\rm{and}\,3.03$, respectively. Particles are stabilized with negatively charged sulfate groups (Microparticles GmbH) in a density-matched 26.3 wt\% aqueous solution of glycerine, with a density $\rho=1062\,$kg/m$^3$ and a viscosity $\mu=2.1\,$mPa.s \citep{Volk2018density}. The charged sulfate groups confer them a small negative surface potential (on the order of $-50$ mV) but sufficient to prevent both their agglomeration and their adhesion to the channel walls \citep{sendekie2016colloidal}. The suspension is prepared with a particle volume fraction of about $2\%$, then inserted in the device and driven downstream the constriction towards a filter which only allows only the fluid to flow through. Particles are therefore initially concentrated in that position.

\begin{figure*}
\centerline{\includegraphics[width=1\textwidth]{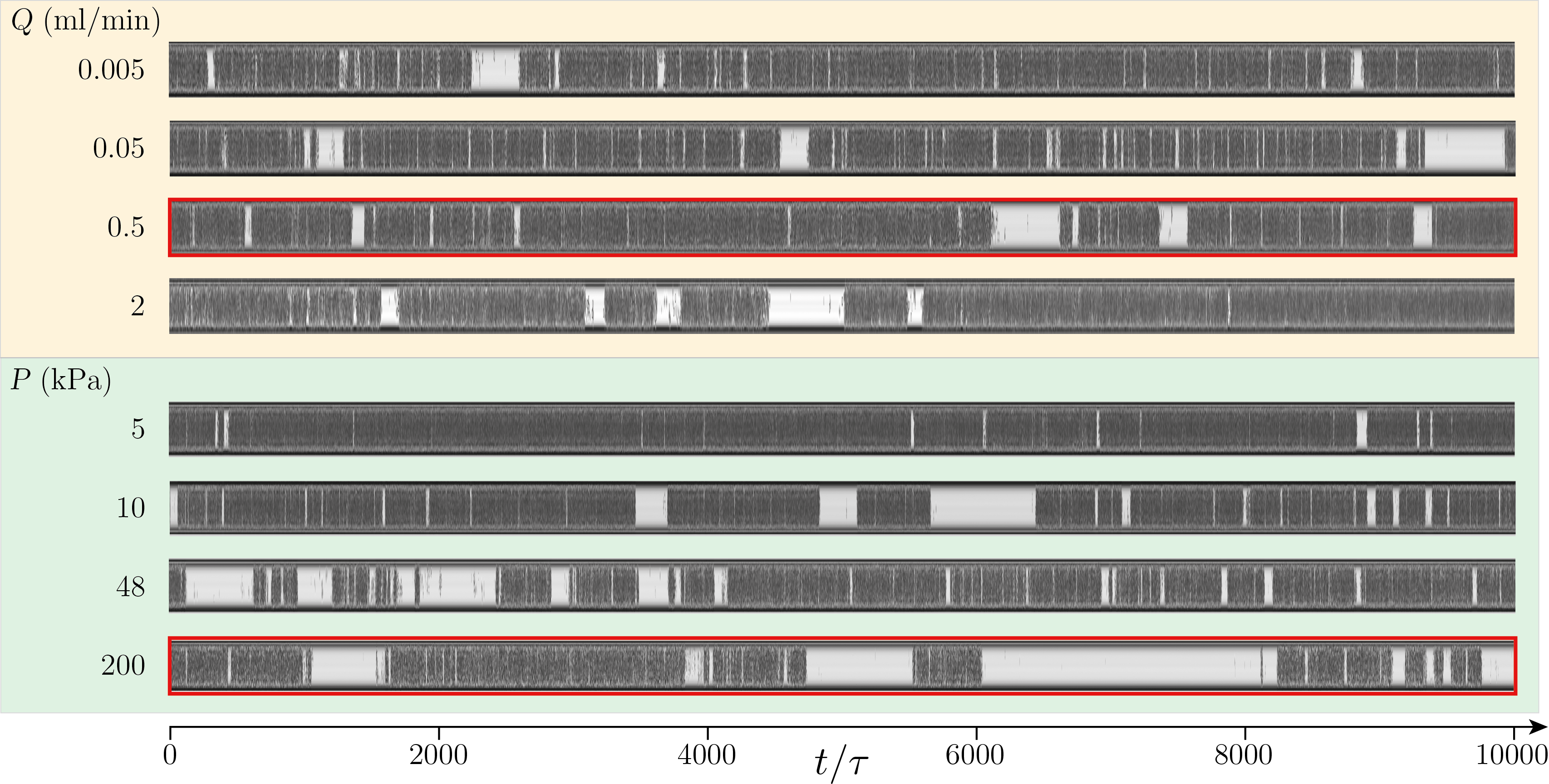}}
  \caption{Spatio-temporal diagrams at the constriction neck for $D/d=4.61$. The top (orange) panel shows that for volume-controlled driving, varying the flow rate $0.005<Q<2$ml/min does not affect the intermittent flow of particles. The bottom (green) panel highlights that for pressure-controlled driving, increasing $P$ induces fewer particles escaping (in dark) before a clog form, and longer arrest times (in grey). The red-framed diagrams highlight two similar $Re$ configurations ({see also Table \ref{fig:table2}}).}
  \label{fig:SpatioTemp}
\end{figure*}
The protocol is similar to the one used in \citet{Souzy2020}: an experiment starts when the flow is reversed and particles are dragged by the fluid towards the constriction. The particle volume fraction as the particle reach the constriction is $\approx60\%$, and the suspension is imaged with a high-speed CMOS camera (PCO.dimax CS1) coupled to an inverted microscope (Nikon Instruments, Eclipse TE2000-U). Figure \ref{fig:setup} presents successive snapshots of a typical intermittent flow experience. The flow may become interrupted by the spontaneous formation of arches spanning the bottleneck (left panel). Although the particle {flow} is abruptly interrupted, the fluid keeps passing through the particles interstices, perturbing the arches which may eventually collapse \citep{Cates1998jamming}. If this happens, the flow of particles is resumed (middle panel) and a burst of particles escape until the development of a new clog arrests the flow again (right panel). Note that arches are stabilized by the mutual friction among particles \citep{sendekie2016colloidal, Wyss2006, Agbangla2014collective}, implying that, despite the presence of electrostatic stabilization, particle-to-particle contact is unavoidable. However, these contacts are not long-lasting as particles separate downstream the constriction upon resuming the flow.

Before discussing about the driving force inducing the flow, we would like to assert two major comments regarding this experimental set-up:
\textit{(i)} the sole possible clogging mechanism here is by {arching}, when particles arriving simultaneously at the constriction form an arch which interrupt the flow of particles, \emph{i.e.} competition of too many particles for little space \citep{Dressaire2017, sendekie2016colloidal}. {Therefore, clogging can not occur by aggregation \citep{Delouche2021}, successive deposition \citep{Agbangla2012, Delouche2020, Dersoir2015clogging, duru2015three}) or sieving ($i.e.$ particles being larger than the constriction \citep{Sauret2014}).}
\textit{(ii)} the reason for using such particle size range is dual: on the one hand, we avoid colloidal particle interactions and Brownian motion as the P\'eclet number $Pe=\langle U\rangle d / D_0> 10^6$, with $D_0=k_BT/3\pi\mu d$ where $k_B$ is the Boltzmann constant, and $T=298K$ is the room temperature. On the other hand, increasing the particle size also involves the flow control of larger volumes of fluid, larger Reynolds number $Re$ and higher working pressures. Therefore, the range of particle size chosen allows us to work with highly monodisperse particles interacting mainly by {hydrodynamic} interactions and low-pressure solid contacts, manipulated via microfluidic technology, which allows us to obtain a high degree of control and reproducibility difficult to achieve at other length scales.

\subsection{Controlling the liquid driving}\label{drivingforce}
To investigate the role of the liquid driving on the clogging/unclogging of constricted particle suspensions, we performed two sets of experiments, where we drive the flow imposing either the flow rate (volume-controlled), either the pressure (pressure-controlled). In both cases, we report the results obtained for neck-to-particle size ratios $D/d= 5.26, 4.61,\,\rm{and}\,3.03$. These specific values of the neck-to-particle size ratios are selected as they allow to explore the limits of aspect ratios for which intermittency has been reported in previous study performed within the same setup \citep{Souzy2020}.

\subsubsection{Volume-controlled driving}\label{FlowRateConfig}

In the first set of experiments, the suspension is driven through the constriction at a constant volume flux using a syringe pump (Harvard Apparatus). The flow rate $Q$ is varied within the range $0.005\!\leq Q\!\leq 2$ ml/min, and the average particle velocity at the constriction neck ranges from $3.9\!\leq\! \langle U\rangle\!\leq\!181$ mm/s, corresponding to Reynolds numbers $Re=\rho \langle U\rangle d /\mu$ in the range $0.042\leq Re\leq 1.99$.  {Note that, even at the highest pressures, the flow is always viscosity-dominated. Reynolds numbers are employed here simply as a way to non-dimensionalise velocities, since the only time scales in the system are given by the driven flow and the liquid's viscosity}. The average particle velocity $\langle U \rangle$ is used to define the Stokes time $\tau=d/\langle U\rangle$ (the time a particle takes to travel its own diameter), which will be used as the characteristic time scale. The use of a syringe pump implies that a constant liquid volume flux is forced through the channel, even when an arch forms at the bottleneck blocking particles. In that case, only liquid flows through the bottleneck and pressure increases due to the increased hydrodynamic resistance. {For safety, the upstream pressure is measured and the experiment is stopped and re-initialized to avoid damage in the setup if the pressure reaches 700 kPa.}

\subsubsection{Pressure-controlled driving}\label{PressureConfig}
The second set of experiments is performed driving the flow with a constant pressure using a microfluidic flow control system (Fluigent). The pressure $P$ is varied within the range $2\!\leq\! P\!\leq\!200$ kPa, and the average particle velocity at the constriction neck ranges from $0.8\!\leq\! \langle U\rangle\!\leq\!17.9$ mm/s, corresponding to Reynolds number $0.009\leq Re\leq 0.198$. Driving the flow with a constant pressure implies that when a clog forms, the hydraulic resistance rapidly increases as the particles accumulate upstream the constriction neck \citep{Kim2017clogging, Sauret2018growth}. Constant pressure configuration can thus be considered as self-regulated: the clog reduces both the liquid and particle flow.


\begin{figure*}
    \centerline{\includegraphics[width=0.8\textwidth]{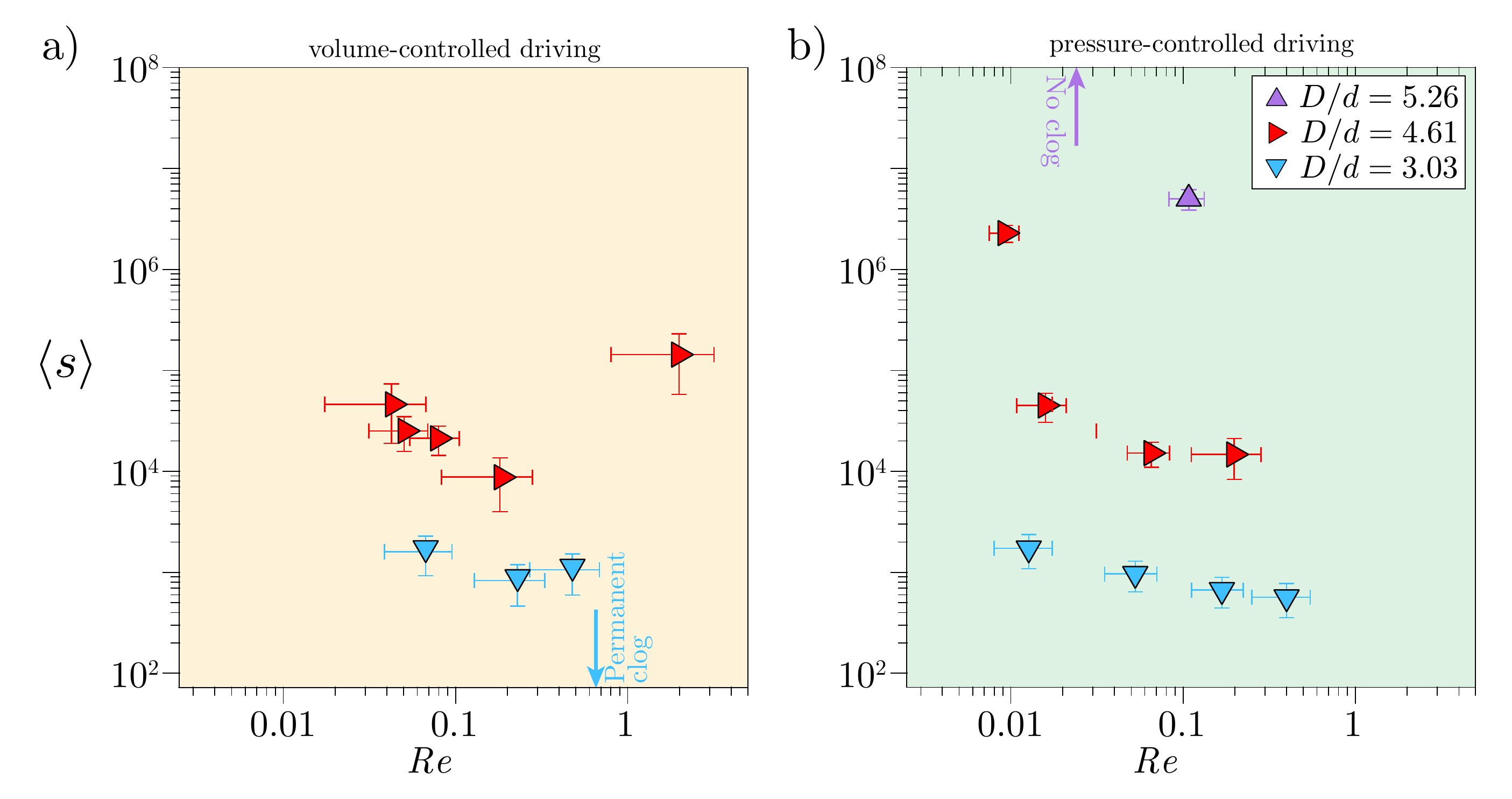}}
    \caption{Average number $\langle s\rangle$ of escapees per burst, for various neck-to-particle size ratios $D/d$. a) In the volume-driven configuration, $\langle s\rangle$ remains almost constant when varying the flow rate $Q$. For $D/d=3.03$, no intermittency is observed for $Re\!>\!0.478$ ($Q\!>\!0.1$ml/min) as any developed clog is permanent and everlasting. b) In the pressure-driven configuration, $\langle s\rangle$ decreases when increasing the imposed pressure $P$. For $D/d=5.26$, the flow of particles is continuous for $Re\leq0.025$ ($P\leq48$kPa).}
    \label{fig:escapees}
\end{figure*}
\section{Results}\label{res}

\normalsize
The overall intermittent behaviour, in which several flow and arrested periods of time alternate, can be better visualized in spatio-temporal diagrams as shown in Figure \ref{fig:SpatioTemp}\footnote{As a guide to the eyes, in the figures an orange background refer to the volume-controlled configuration while a green background refer to the pressure driven configuration.}. There, in order to analyze the different regimes of particle flow, we report the results obtained for $D/d = 4.61 $. A spatio-temporal diagram is constructed by selecting a vertical line of unit pixel width at the middle of the constriction. This line is then sampled for every frame and is stacked alongside. 
Interestingly, for the volume-controlled driving (top orange panels), there is no clear qualitative difference in the flow behaviour in the investigated range of flow rate. Conversely, for the pressure-controlled driving (bottom green panels), the diagrams clearly reveal a difference in the flow behaviour: for $P=5$ kPa, the flow of particles (which appears in dark) is almost continuous, with only few short-lasting arrest times (which appears in light grey). As $P$ is increased up to $P=200$ kPa, less particles escape when a burst of particles is released, and the arrest times become longer and more abundant.

In what follows, we will quantify the intermittent dynamics by analyzing separately the {arch} formation and destruction processes by looking at the statistics of burst sizes and arrest times respectively.

\subsection{{Arch formation/clogging probability}}\label{Clog_dev}

Similarly to pedestrians \citep{Garcimartin2016pedestrians}, animal flocks \citep{Garcimartin2015} and avalanches  \citep{Fisher1998}, the number of entities escaping per burst has been found to follow an exponential distribution in constricted flow of suspensions \citep{Souzy2020}. {This reveals that the arch formation follows a Poissonian process, with an exponential dependence of the discharged particle mass with the neck-to-particle size ratio.} Consequently, one can estimate the clogging probability by estimating the average number $\langle s\rangle$ of escaping particles per burst. 

Given the discrete nature of the system, defining an arrest time threshold to set apart successive bursts is not straightforward. We define a clog event as any event where the time lapse $T$ between the passage of two consecutive particles is such that $T\geq10\tau$, where $\tau$ is the characteristic Stokes time. We checked that this arbitrary choice does not affect the general trend of our results.

Figure \ref{fig:escapees} shows $\langle s\rangle$ for various neck-to-particle size ratios $D/d$. The error bars stand for the standard deviation of the flowing particle velocity measurement $U$, obtained by performing Particle Image Velocimetry at the constriction neck when particles are flowing. More than 1500 bursts have been analysed in each case. 
{As it is well-known in dry systems \citep{to2001jamming,zuriguel2005jamming,Thomas2015}, the neck-to-particle size ratio is the governing parameter for the intermittent clogging statistics, therefore it is not surprising to observe a steep increase of $\langle s \rangle$ for higher $D/d$.}

\begin{figure*}
    \centerline{\includegraphics[width=0.6\textwidth]{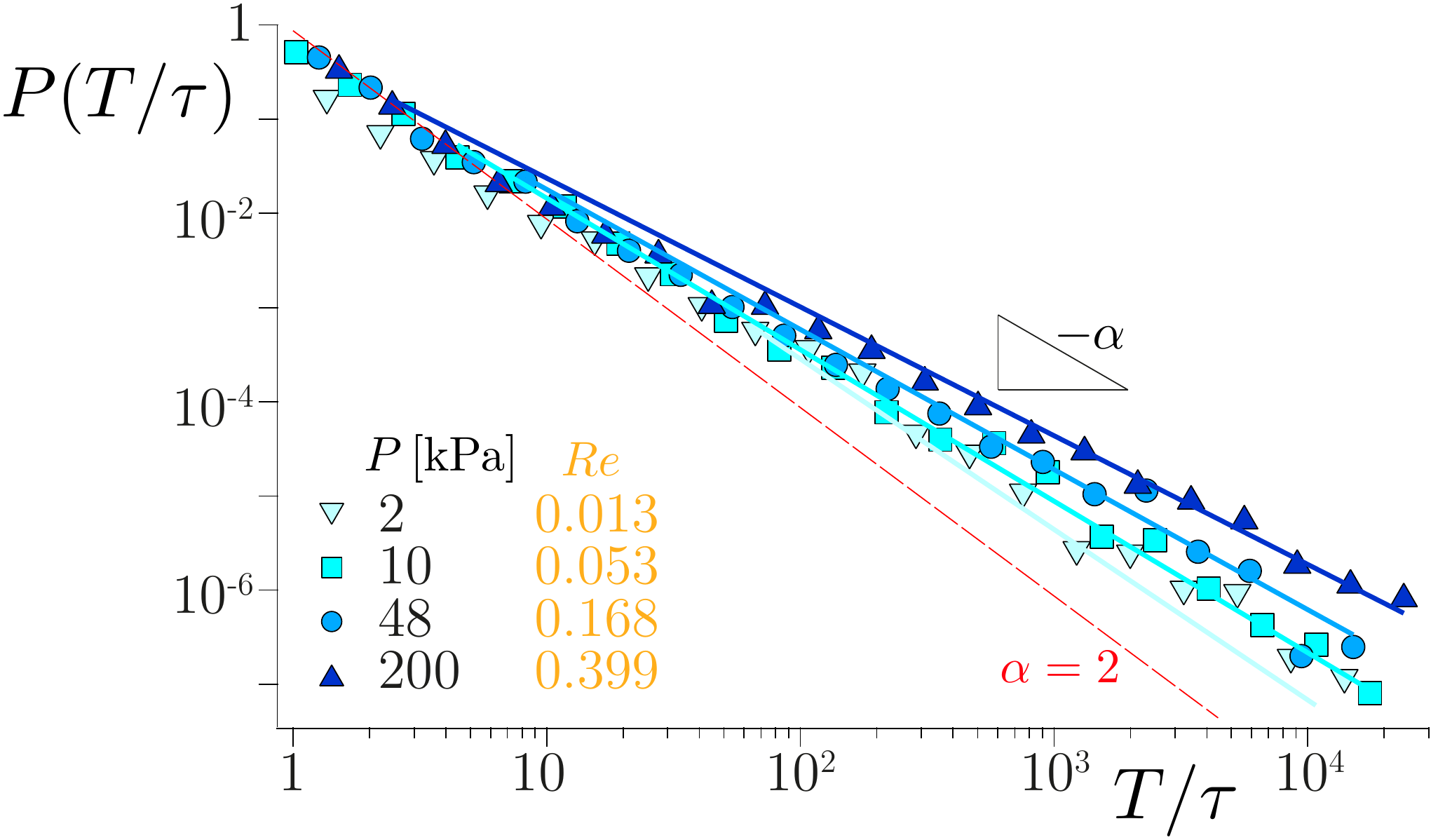}}
    \caption{Distribution of the arrest time lapses normalized by the Stokes time $T/\tau$ for $D/d=3.03$ with pressure controlled driving. The lines correspond to the best power law fits with their exponent $\alpha$, as given by the Clauset-Shalizi-Newman method \citep{Clauset2009}.}
    \label{fig:P(T)}
\end{figure*}
{For the volume-controlled configuration (Figure \ref{fig:escapees}a, left orange panel), we can first observe that varying the flow rate $Q$ has a minor effect on the clogging probability, as $\langle s\rangle$ remains almost constant over the investigated range of $Q$. The average number of escapees $\langle s\rangle$ increases one or two orders of magnitude when we reduce the particle size by just {33\%} (from $D/d=3.03$ to 4.61). An even smaller relative decrease of the particle size results in a continuous flow of particles for $D/d=5.26$, in which $\langle s\rangle$ could not be defined (over the range of investigated $Q$), which is consistent with what was reported in \citet{Souzy2020}.
On the other side of the parametric range of this intermittent regime, for larger particle sizes, no intermittency is observed for $D/d=3.03$ after a critical Reynolds number/flow rate ($Q\geq0.5$ ml/min). After this critical point, any particle arch is either permanent or lasting more than our largest measurable timescale. This is limited by the maximum time that can be captured on our camera's digital memory, which sets the maximum number of frames that we can capture on each experiment. Nonetheless, the experiments are designed to guarantee that the duration of each movie is at least four orders of magnitude larger than the Stokes time \footnote{The maximum number of frames we can acquire is $2^{16}=65536$. For this particular experiment ($D/d=3.03$, flow rate $Q=0.5$ mL/min), running the camera at 5000 fps sets the maximum recording time to 13 sec, which represents roughly more than 10 000 stokes time $\tau$ based on the flow of particles that manage to get through the constriction prior to the eventual formation of a `permanent' arch.}.
}

For the pressure-controlled configuration (Figure \ref{fig:escapees}b, right green panel), the trend is noticeably different. A pressure increase $P$ (and therefore in particle velocity and $Re$) clearly leads to a decrease in the average number of escapees $\langle s \rangle$, thus increasing the clogging probability. Note that this is in direct contradiction with results reported for clogging with cohesive suspensions, in which clogs are built up by successive deposition. In such cases, higher pressure leads to increased shear and to the detachment of deposited particles, and thus to a decrease of the clogging probability \citep{Delouche2021, Dersoir2015clogging, Kim2017clogging}. 
It is nonetheless a striking result even in the absence of cohesive forces, as it reveals a fundamental and yet unexpected effect of the driving force on the clog development process for constricted particle suspensions. 
This result can be observed for instance by comparing the case of $D/d=4.61$ under different driving. For comparable particle velocities ($Re=0.181$, in volume-controlled, and $Re=0.198$, in pressure-controlled, highlighted with red-frames both in Figure \ref{fig:SpatioTemp} and {Table \ref{fig:table2}), we obtain 10 times more escapees for a volume-controlled than with the pressure-controlled system. 
The same trend is observed for the other data sets in different manners. At first glance, the values of $\langle s \rangle$ for the largest particle size $D/d=3.03$ look comparable for both driving methods. However, the pressure-driven set has a clearer inverse dependency with increased pressure. Interestingly, the intermittent regime remains stable up to $Re=0.4$ in the pressure-driven case, while it clogs permanently for the volume-controlled case. 
Another remarkable example of such a trend between imposing the flow rate and the pressure is given for the smallest particle size, corresponding to $D/d=5.26$: for comparable particle velocities, we observe a continuous particle flow when the flow rate is imposed, while the flow is intermittent for the pressure-driven configuration.

It has been previously suggested that clogging in constricted suspensions by mechanical forces occurs when the particle rate reaches certain critical value, below which bridging will not occur since hydrodynamic forces at the constriction have to overcome the repulsion barrier \citep{Agbangla2012, hsu2021roughness}. Our results reveal that clogging by arching of non-cohesive suspensions is not only set by the particle velocity \citep{Ramachandran1999, Guariguata2012} but also by the liquid driving choice, as for a given aspect ratio and particle velocity, different qualitative and quantitative behaviour is observed depending on whether the flow is driven imposing a flow rate or pressure.

While the dependence of the arch formation on the liquid driving method is kind of subtle in the data shown so far, it will be much more noticeable on the arch destruction process, which will be analysed using the statistical distribution of time-lapses.


\begin{figure*}
   \centerline{ \includegraphics[width=0.8\textwidth]{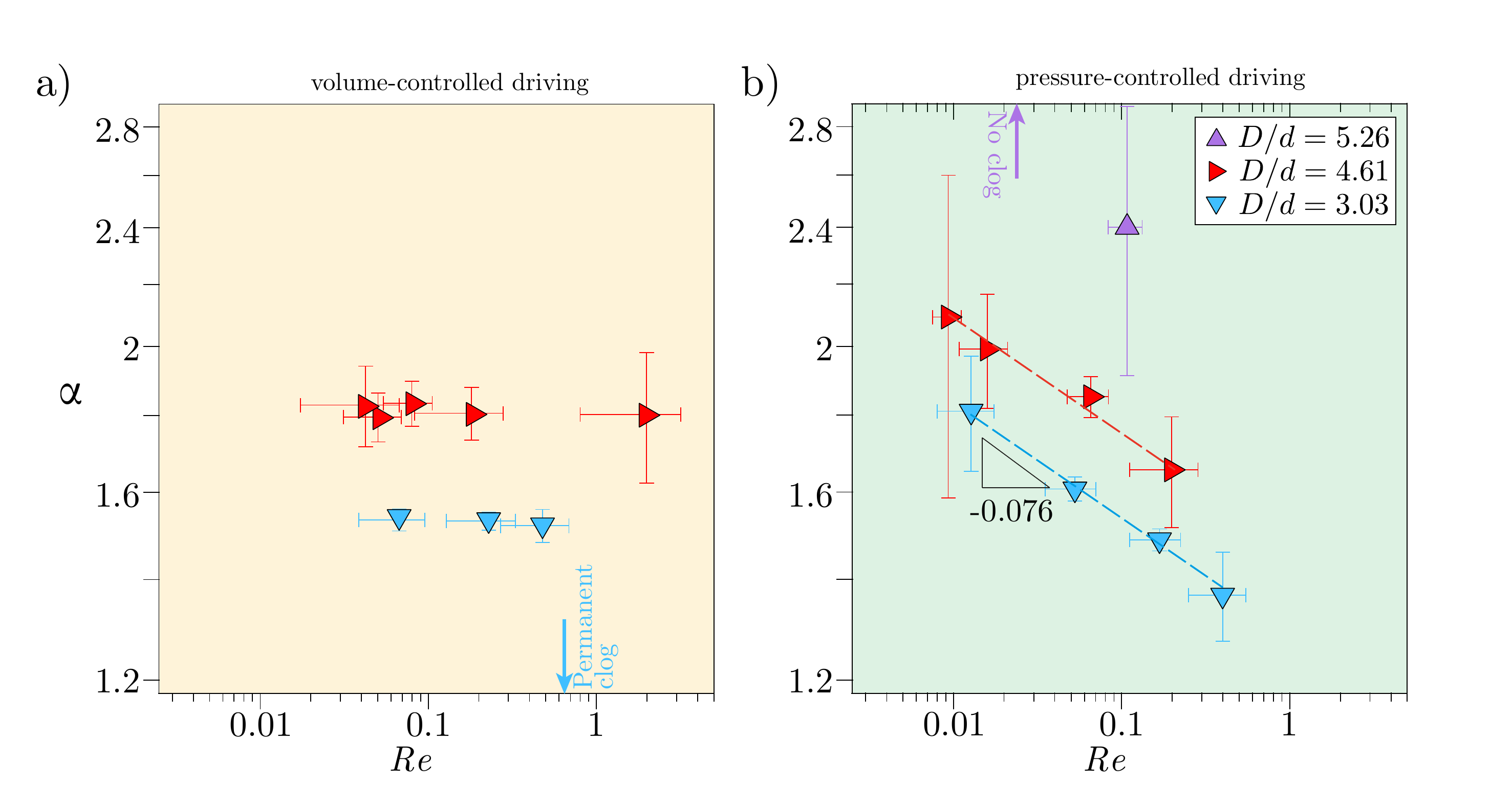}}
    \caption{Exponent $\alpha$ of the power-law tail of the timelapse distribution $P(T/\tau)\propto(T/\tau)^{-\alpha}$. a) In the volume-driven configuration, $\alpha$ remains almost constant when varying the flow rate $Q$. For $D/d=3.03$, no value of $\alpha$ could be defined for $Re\!>\!0.478$ ($Q\!>\!0.1$ml/min) as there is no intermittency. b) In the pressure-driven configuration, $\alpha$ exhibit a power-law decrease $\alpha\propto Re^{-0.076}$. For $D/d=5.26$, no value of $\alpha$ could be defined for $Re\leq0.025$ ($P\leq48$kPa) as there is a continuous flow of particle. The explicit values of $\alpha$ can be found in { Table \ref{fig:table2} }.}
    \label{fig:alpha}
\end{figure*}

\subsection{{Arch destruction/unclogging probability}}\label{Clog_destr}

In order to investigate the unclogging process, we now turn to analyse the probability distributions of time lapses $T$ between the passage of consecutive particles. This approach has been extensively implemented in previous studies on intermittent flowing systems, such as hungry sheep herds \citep{Garcimartin2015}, pedestrian crowds \citep{Helbing2005, Krausz2012,Garcimartin2016pedestrians}, mice escaping a water pool \citep{Saloma2003}, or vibrated silos of dry granular material \citep{Janda2009, Lastakowski2015}. In such systems, the distribution of arrested time lapses exhibits a power-law tail $P(T)\propto T^{-\alpha}$, a signature of systems susceptible of clogging \citep{Zuriguel2014,Zuriguel2020}. Furthermore, the value of the exponent $\alpha$ can be directly related to the long-term behaviour of the system: the average time lapses $\langle T\rangle$ can only be defined for distributions fulfilling $\alpha>2$, while $\langle T\rangle$ diverges for $\alpha\leq2$. This feature has therefore been interpreted as a transition to a scenario in which a permanent clog could eventually occur with finite probability. Indeed, for $\alpha\leq2$, there is a non-zero probability of observing everlasting clogs, while for $\alpha>2$, the system can be temporary blocked due to the formation of clogs but no arch will persist infinitely. More detailed discussions can be found in \citet{Zuriguel2014}, \citet{Zuriguel2020}, or in \citet{Garcimartin2021}.

{Figure \ref{fig:P(T)} presents the probability distribution of the arrest lapses obtained for $D/d=3.03$ with pressure-controlled driving. The distribution $P(T/\tau)$ exhibits the characteristic power-law tail $P(T/\tau)\propto(T/\tau)^{-\alpha}$. The time lapse probability distribution is obtained for  $0.013\leq Re\leq 0.399$ (corresponding to $2\leq P\leq200$ kPa)
. Note that we have been able to measure time lapses up to four orders of magnitude larger than the Stokes time, and that around 2000 bursts have been analyzed for each investigated configuration.}
Finding the right parameters for power-law tails can easily suffer from arbitrary biases, therefore the exponent $\alpha$ of the power-law tail is obtained using the rigorous and widely accepted Clauset-Shalizi-Newman method \citep{Clauset2009}, which also yields the estimated error of the fit.

Figure \ref{fig:alpha} presents the value of $\alpha$, for various neck-to-particle size ratios $D/d$ and different liquid driving methods. The first thing to notice,  as previously reported \citep{Souzy2020}, is that the value of $\alpha$ decreases significantly with the neck-to-particle size ratios $D/d$. This is expected: the smaller the particles, the higher the value of $\alpha$, thus the higher is the probability of unclogging. This highlights the fact that arches composed of more particles are less stable, and thus more prone to break due to the perturbations induced by the interstitial flow. In other words, shorter arches are stronger than longer ones.

The response of the system to different  driving liquid methods is here more noticeable: for the volume-controlled configuration (Figure \ref{fig:alpha}a, left orange panel), the exponent $\alpha$ remains constant over more than two decades of flow rate values, while there is a clear and robust dependence for the pressure-controlled configuration (Figure \ref{fig:alpha}b, right green panel). More specifically, in the pressure-controlled configuration, $\alpha$ exhibits a clear power-law decrease which is well fitted by $\alpha\propto~Re^{-0.076}$ over two decades. This trend is found consistently both for $D/d=3.03$ and $4.61$. The decrease in $\alpha$ indicates that, for higher imposed pressure, the probability of breaking an arch decreases. Again, note that this is precisely the opposite trend observed with cohesive suspensions \citep{Delouche2021, Dersoir2015clogging, Kim2017clogging}. As mentioned in the previous section, for $D/d=5.26$ the flow is found to be intermittent only for $Re\geq0.108$ ($P\geq200$ kPa), thus the exponent $\alpha$ could only be determined for the maximum experimental pressure we could impose.


\begin{table*}
  \begin{center}
\def~{\hphantom{0}}
  \centerline{\includegraphics[width=0.5\textwidth]{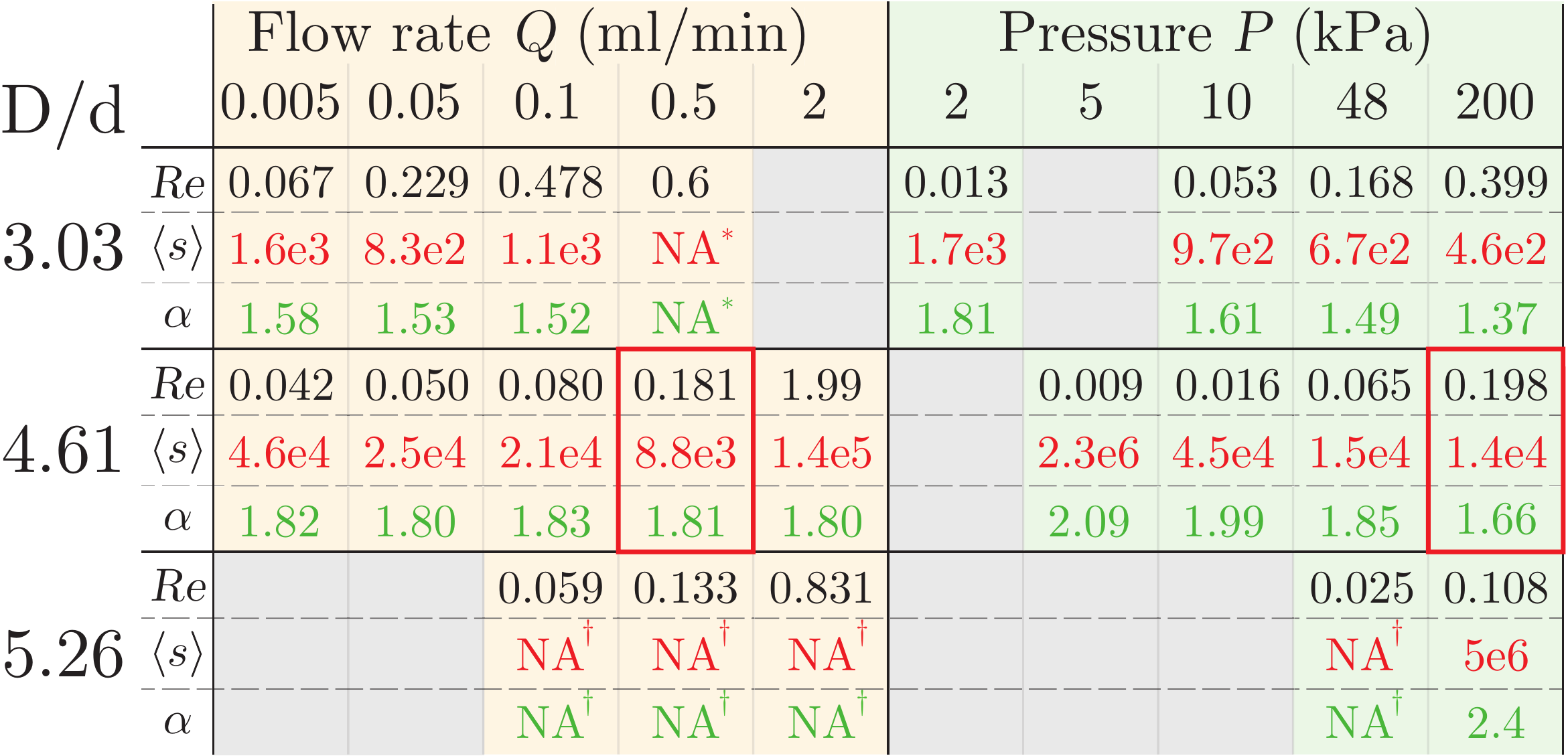}}

\caption{Values of $Re$, $\langle s\rangle$ and $\alpha$ for various neck-to-particle size ratios, for volume-controlled and pressure-controlled driving. Grey areas stand for non-investigated parameters values. NA stand for non-intermittent configurations for which we could not estimate the corresponding values: $^\dagger$ continuous particle flow, and $^*$ everlasting clog. Red frames highlight two similar Reynolds number configurations.}
\label{fig:table2}
\end{center}
\end{table*}
\section{Discussions \& Conclusion}\label{concl}


Using direct visualization at the particle scale, flow measurements and statistical techniques, we investigate the role of the liquid driving in the intermittent flow of non-cohesive particle suspension through a constriction by reporting the average number of escapees per burst $\langle s\rangle$ (which reflects the probability of clogging), and the exponent $\alpha$ of the power-law tail of the distribution of arrest time (which reflects the probability of breaking a clog).

Our aim was to answer the apparently trivial question: \textbf{will more particles pass through the constriction if we ``push'' harder?} The answer we have found is non-trivial and it depends on the way the liquid is being driven through the channel. Our results reveal striking discrepancies for both the clog development process and the clog destruction process, depending on whether the flow rate or the pressure is imposed.

When a constant \emph{liquid flow rate} is imposed, both $\langle s\rangle$ and $\alpha$ appear to be constant over the range $0.042\leq Re\leq 1.99$ ($0.005\leq Q\leq 2$ ml/min), the statistics of forming or breaking an arch are practically independent on the flow rate imposed.
There is however an exception for the largest particle size tested ($D/d=3.03$), for which we observe a sudden transition from intermittent to permanent clog when the $Re$ is increased from 0.48 to 0.6 (\ref{fig:escapees}a and \ref{fig:alpha}a). We do not have a clear explanation to this sudden transition. It could be due to certain critical behaviour at a certain flow rate, but it could also be an experimental limitation due to our limited time to capture long enough arrest times.
Note that these results are in agreement with our previous work \citep{Souzy2020}, performed in volume-controlled configuration, for which we only found intermittent regimes with $\alpha\leq2$. Consequently, our previous and current results imply that volume-control leads to intermittent regimes with high probability of persistent clogs.

For the \emph{pressure-controlled} configuration, both $\langle s\rangle$ and $\alpha$ decrease when increasing the pressure $P$. Such trend is found consistently for $D/d=3.03$ and $4.61$. For the smallest particle size ($D/d=5.26$), the system flows continuously for low pressures, and it becomes intermittent after increasing the pressure (or $Re$) above certain value ($P\geq200$kPa, $Re=0.108$). Remarkably, for $D/d=4.61$ and $5.26$, values of $\alpha>2$ were found in the pressure-controlled configuration, thus indicating that the intermittent regimes would continue indefinitely, with clogs that would not persist. 
Thus, in the pressure-controlled configuration, a non-cohesive suspension passing through a constriction behaves similarly to other scenarios where clogging transitions have been reported based on the power-law tails of the arrest times, like vibrated silos \citep{Janda2009}, Brownian particles \citep{Hidalgo2018}, pedestrians \citep{Zuriguel2014}, and self-propelled robots \citep{Patterson}. All those systems reported a fairly smooth transition from an intermittent clogged state ($\alpha\leq2$) to a continuous flow, passing through a region of intermittent flow with $\alpha>2$.

\begin{table*}
  \begin{center}
\def~{\hphantom{0}}
\begin{tabular}{ll|lll}
             & $\uparrow$ Gravity \citep{Arevalo2014}    & $\uparrow$ Vibrations \citep{Mankoc2009, Janda2009}     &  &  \\
Dry granular & \begin{tabular}[c]{@{}l@{}}\textcolor{ForestGreen}{$\uparrow \langle s\rangle$},  \textcolor{Red}{$\downarrow P(\rm{clog~formation})$}\\ \textcolor{Red}{$\downarrow \alpha$},  \textcolor{Red}{$\downarrow P(\rm{breaking\,clog})$}\end{tabular} & \begin{tabular}[c]{@{}l@{}}\textcolor{ForestGreen}{$\uparrow \langle s\rangle$},  \textcolor{Red}{$\downarrow P(\rm{clog~formation})$}\\ \textcolor{ForestGreen}{$\uparrow \alpha$},  \textcolor{ForestGreen}{$\uparrow P(\rm{breaking\,clog})$}\end{tabular}&  &  \\ \cline{2-3}
             &  $\uparrow$ Driving pressure                                                                                                                           & $\uparrow$ Temperature \citep{Hidalgo2018}&  &  \\
Non-cohesive suspension  &  \begin{tabular}[c]{@{}l@{}}\textcolor{Red}{$\downarrow \langle s\rangle$},  \textcolor{ForestGreen}{$\uparrow P(\rm{clog~formation})$}
\\ \textcolor{Red}{$\downarrow \alpha$},  \textcolor{Red}{$\downarrow P(\rm{breaking\,clog})$} 
\end{tabular}& \begin{tabular}[c]{@{}l@{}}\textcolor{Red}{$\downarrow \langle s\rangle$},  \textcolor{ForestGreen}{$\uparrow P(\rm{clog~formation})$}
\\ \textcolor{ForestGreen}{$\uparrow \alpha$},  \textcolor{ForestGreen}{$\uparrow P(\rm{breaking\,clog})$}\end{tabular} &  & \end{tabular}
\caption{Strategies leading to a modification of the number $\langle s\rangle$ of escapees released per burst and of the exponent $\alpha$ of the power law fits of the time lapses distribution.}
\label{tab:Effects}
\end{center}
\end{table*}
It is unclear for us why such a smooth transition does not occur with a volume-controlled driving.
At this respect, note that due to the narrow size distribution of our particles, the number of particle configurations that result in a stable arch is greatly reduced. That might have an impact in the different regimes that could manifest, and could be an explanation for the sudden transition from an intermittent clogged state ($\alpha\leq2$) to a continuous flow regime in the volume-controlled driving. Nevertheless, whether such regime ($\alpha>2$) exists or not in the volume-controlled driving configuration remains an open question, which we will continue studying in the future. 

We could rationalize the results in the following way.
On the one hand, the arch formation follows generally similar trends for both liquid driving configurations, and it follows an intuitive trend: a higher particle rate leads to more chances for clog formation, and therefore a decrease of the burst size. However, quantitatively the average burst size does depend on the choice of liquid driving. For a given aspect ratio and particle velocity, different $\left<s\right>$ can be observed depending on whether the flow is driven imposing a flow rate or pressure. On the other hand, the clog breakup seems to be more sensitive to the driving method: The volume-controlled system responds increasing the pressure locally in the region of the arch, and therefore the flow velocity too. Particles progressively accumulate behind the newly formed arch, increasing the hydrodynamic resistance further as the aggregate's length increase. Since the liquid flow rate needs to be kept constant, the liquid velocity, and therefore the drag pushing particles together increases. This would lead to an increased and more stable contact chain force network, probably close to jamming, following Cates's approach \citep{Cates1998jamming}. That compact network imposes such a large hydrodynamic resistance, that a flow rate increase (in the range explored here) does not change significantly the stability of the arch. Consequently, $\alpha$ remains effectively constant for the range of flow rates explored.
However, in a pressure-controlled system, the local pressure, and therefore the liquid velocity decreases in the arch and its vicinity, leaving a rather loose packing in the region behind the arch. An increase in pressure leads to a larger compacted region behind arch stabilizing it. Therefore, lower pressure results actually in less stable archs, which translates into better particle transport: a fluid-dynamical version of the ``slower is faster'' effect observed in pedestrians \citep{Zuriguel2014}.

Note that these results stand in strike contrast with what is reported for \emph{cohesive} particle suspensions, which follows a more intuitive trend. The shear flow counteracts particle adhesive forces, and therefore an increase of both pressure or flow rate leads a decrease of the clogging probability \citep{Delouche2021, Dersoir2015clogging, Kim2017clogging}. 

Our results provide new efficient strategies for controlling the flow of non-cohesive particle suspensions driven through constrictions. Several strategies have already been investigated based on different parameters (see summary in Table \ref{tab:Effects}). For example, \citet{Arevalo2014} showed using simulations that gravity decreases both the probability of clogging and of unclogging for granular inert particles passing through a bottleneck. \citet{Janda2009} and \citet{Mankoc2009} both showed experimentally that for dry granular silos, applying strong vibrations decreases the clogging probability, and that the remaining clogs break also more easily. We herein show that, for non-Brownian and non-cohesive suspensions, increasing the driving pressure leads to an increase of the clogging probability and a decrease of the unclogging probability. Consequently, an efficient strategy for limiting clogging of non-cohesive particle suspension is to drive the flow imposing lower pressure. 



\section{Acknowledgements}
This work was supported by the ERC (European Research Council), Starting Grant (grant agreement No.678573). The authors would like to acknowledge the motivation for this work and insights from Iker Zuriguel. 






\bibliographystyle{apsrev4-1}
\bibliography{Bibliography}

\end{document}